\documentclass[12pt]{article}
\usepackage{amssymb}
\makeatletter
\def\numberbysection{\@addtoreset{equation}{section}
 	\def\theequation{\thesection.\arabic{equation}}}
\makeatother

\numberbysection

\newcommand{\dmat}[3]{%
    \begin{array}{ccc}
    	#1 &  &   \\
    	 & #2 &   \\
    	 &  & #3
    \end{array}}
\newcommand{\umat}[3]{%
    \begin{array}{ccc}
    	 & #1 & #2   \\
    	 &  & #3  \\
    	 &  & 
    \end{array}}
\newcommand{\lmat}[3]{%
    \begin{array}{ccc}
    	 &  &   \\
    	#1 &  &   \\
    	#2 & #3 & 
    \end{array}}
\newcommand{\be}{\begin{eqnarray}}
\newcommand{\ee}{\end{eqnarray}}
\newcommand{\non}{\nonumber}
\newcommand{\id}{\mathbb{I}}

\begin{document}

\begin{titlepage}
\strut\hfill UMTG--237
\vspace{.5in}
\begin{center}

\LARGE Boundary quantum group generators of type A\\[1.0in]
\large Rafael I. Nepomechie\\[0.8in]
\large Physics Department, P.O. Box 248046, University of Miami\\[0.2in]  
\large Coral Gables, FL 33124 USA\\

\end{center}

\vspace{.5in}

\begin{abstract}
We construct boundary quantum group generators which, through linear
intertwining relations, determine nondiagonal solutions of the
boundary Yang-Baxter equation for the cases $A^{(1)}_{n-1}$ and
$A^{(2)}_{2}$.
\end{abstract}

\end{titlepage}

\setcounter{footnote}{0}

\section{Introduction}\label{sec:intro}

An effective means of finding solutions $R(u)$ of the Yang-Baxter 
equation \cite{ZZ}-\cite{KBI}
\be
R_{12}(u-v)\ R_{13}(u)\ R_{23}(v)
= R_{23}(v)\ R_{13}(u)\ R_{12}(u-v) 
\label{YB}
\ee 
is the so-called quantum group approach \cite{KR, Ji}, which reduces
the problem to a linear one.  Indeed, $R$ matrices corresponding to
vector representations of all non-exceptional affine Lie algebras were
determined in this way in \cite{Ji}.

A similar approach is clearly desirable for finding solutions $K(u)$
of the boundary Yang-Baxter equation \cite{Ch, Sk, GZ}
\be
R_{12}(u-v)\ K_{1}(u)\ R_{21}(u+v)\ K_{2}(v)
= K_{2}(v)\ R_{12}(u+v)\ K_{1}(u)\ R_{21}(u-v) \,.
\label{boundaryYB}
\ee 
With this goal in mind, the study of boundary quantum groups was
initiated in \cite{MN1}.  In particular, for the case that the $R$
matrix corresponds to the spin ${1\over 2}$ representation of
$A^{(1)}_{1}$, two matrices $Q_{0}(u)$, $Q_{1}(u)$ were constructed
which determine (up to an overall unitarization factor which does not
concern us here) the $K$ matrix \cite{GZ, dVGR} through the linear
``intertwining'' relations
\be
K(u)\ Q_{j}(u) = Q_{j}(-u)\ K(u) \,, 
\label{intertwining}
\ee 
where here $j = 0 \,, 1$.  This approach has recently been generalized
in \cite{DM} to the $A^{(1)}_{n-1}$ case where vector solitons are
reflected into solitons in the conjugate vector representation
\cite{Ga}.  Moreover, this boundary quantum group approach has been
used in \cite{DN} to determine $K$ matrices for higher representations
of $A^{(1)}_{1}$.

Since the boundary quantum group ``generators'' $Q_{j}(u)$ determine
(through the intertwining relations (\ref{intertwining})) solutions
$K(u)$ of the boundary Yang-Baxter equation, they seem to be important
objects. However, very little is yet known about them.

In this Letter, we construct such boundary quantum group generators for
the $A^{(1)}_{n-1}$ case where vector solitons are reflected into vector
solitons (i.e., not into their conjugates) , as well as for the
$A^{(2)}_{2}$ case.  The corresponding nondiagonal $K$ matrices
\cite{AR} and \cite{LS, Ki} are generalizations of diagonal $K$ matrices
which were found earlier in \cite{dVGR} and \cite{MN2}, respectively.
In Section \ref{sec:A1n} we treat the $A^{(1)}_{n-1}$ case, and in 
Section \ref{sec:A22} we discuss the $A^{(2)}_{2}$ case. We end with 
a brief discussion of our results in Section \ref{sec:disc}.

\section{The $A^{(1)}_{n-1}$ case}\label{sec:A1n}

Let us consider the case that the $R$ matrix corresponds to the 
vector representation of $A^{(1)}_{n-1}$, $n \ge 3$. It is given 
by \cite{Ji, BdVV} 
\be
R(u) &=& \sinh({n u\over 2} + \eta) 
\sum_{k=1}^{n} E_{k\,, k}\otimes E_{k\,, k}+
\sinh({n u\over 2}) 
\sum_{k \ne l} E_{k\,, k}\otimes E_{l\,, l} \non \\
&+& \sinh (\eta) \left( e^{n u\over 2}\sum_{k<l} 
+ e^{-{n u\over 2}}\sum_{k>l} \right) E_{k\,, l}\otimes E_{l\,, k} \,,
\label{RA1n}
\ee
where $\eta$ is the anisotropy parameter, and $E_{l\,, k}$ denotes the
elementary $n \times n$ matrix with matrix elements $(E_{l\,,
k})_{\alpha \beta} = \delta_{l \alpha} \delta_{k \beta}$.  We remark
that we work with the $R$ matrix in the so-called homogeneous
gradation.

Consider the set of generators
\be 
Q_{0}(u) &=& e^{n u} E_{n \,, 1} + e^{-n u} E_{1 \,, n} 
+ e^{2 \epsilon} e^{- \sigma (-E_{1 \,, 1} + E_{n \,, n})} \,,
\label{charge0A1n} \\
Q_{k-1} &=& E_{k \,, 1} + e^{\sigma} E_{1 \,, k} 
+ E_{n \,, k} + e^{-\sigma} E_{k \,, n}  
+  e^{2 \pi i (k-1)\over (n-1)} E_{k \,, k} \,, \qquad 
k = 2\,, \ldots \,, n-1 \,,
\label{chargesA1n}
\ee
where $\epsilon$ and $\sigma$ are arbitrary boundary parameters. The 
intertwining relations (\ref{intertwining}) determine the following
$K$ matrix:
\be
K(u) &=& \id - {e^{-2 \epsilon}\over \sinh \sigma} \Bigl[
\sinh( n u) (E_{n \,, 1} + E_{1 \,, n}) +  
\sinh (\sigma) \left( e^{n u} E_{1 \,, 1}+ e^{-n u}  E_{n \,, n} \right)
\non \\
&+& \sinh( n u + \sigma) \sum_{k=2}^{n-1} E_{k\,, k} \Bigr] \,,
\label{KA1n}
\ee 
where $\id$ is the identity matrix.
This is essentially the same solution of the boundary Yang-Baxter 
equation (\ref{boundaryYB}) which was found by Abad and Rios \cite{AR}.
Indeed, the latter solution appears to have more boundary parameters: 
$\rho_{a} \,, \rho_{b}\,, \rho_{c}\,, \rho_{d} \,, \epsilon$, with one 
constraint 
\be
\rho_{c}\rho_{d} = \rho_{b} (\rho_{b} + \rho_{a}  e^{-\epsilon}) \,.
\label{constraint}
\ee  
However, by rescaling the $K$ matrix, one can set $\rho_{b}=1$. By a ``gauge'' 
transformation which leaves the $R$ matrix unchanged, 
\be
R_{12}(u) \mapsto M_{1} M_{2} R_{12}(u) M^{-1}_{1} M^{-1}_{2} = R_{12}(u)\,, 
\qquad
K(u) \mapsto M K(u) M^{-1} \,, 
\ee
with $M = diag(1 \,, 1 \,, \ldots \,, 1 \,, \sqrt{\rho_{d}}/\sqrt{\rho_{c}})$
one can bring
$\rho_{c}$ and $\rho_{d}$ to be equal, $\rho_{c}=\rho_{d} \equiv e^{-\sigma}$.
The constraint (\ref{constraint}) then fixes 
$\rho_{a} = e^{\epsilon} (e^{-2\sigma}-1)$. That is, there are only 
two independent boundary parameters, $\epsilon$ and $\sigma$. 
Finally, it should be noted that Abad and Rios work with the $R$
matrix in the so-called principal gradation, which is related to the 
$R$ matrix in the homogeneous gradation by the gauge transformation
\be
R^{prin}_{12}(u) = M_{1}(u)\ R^{hom}_{12}(u)\ M_{1}(-u) \,,
\ee
where $M(u) = diag(1 \,, e^{u} \,, e^{2u} \,, \ldots)$.
Hence, the $K$ matrices are also related by 
a corresponding transformation \cite{MN3}
\be
K^{prin}(u) = M(u)\ K^{hom}(u)\ M(u) \,.
\ee

We remark that the particular set of diagonal terms $e^{2 \pi i
(k-1)/(n-1)} E_{k \,, k}$ in (\ref{chargesA1n}) is merely one
convenient choice.  Indeed, generic diagonal terms will again lead to
the same $K$ matrix (\ref{KA1n}).

We also emphasize that the solution (\ref{charge0A1n}) - (\ref{KA1n})
has two continuous boundary parameters.  In contrast, for the case
that vector solitons reflect into conjugate vector solitons considered
in \cite{DM, Ga}, there are no continuous boundary parameters.

\section{The $A^{(2)}_{2}$ case}\label{sec:A22}

We now consider the case of the Izergin-Korepin $R$ matrix \cite{IK},
which corresponds to the vector representation of $A^{(2)}_{2}$.  It
can be written in the following form \cite{KS0}, \cite{MN4}
\be
R(u) = \left(
    \begin{array}{c|c|c}
    	\dmat{c}{b}{d} & \lmat{e}{\relax}{g} & \lmat{\relax}{f}{\relax}  \\
		\hline
		\umat{\bar e}{\relax}{\bar g} & \dmat{b}{a}{b} & \lmat{g}{\relax}{e}  \\
		\hline
    	\umat{\relax}{\bar f}{\relax} & \umat{\bar g}{\relax}{\bar e} & 
    	\dmat{d}{b}{c}
    \end{array}
\right)
\ee
where
\be 
a &=& \sinh (u - 3\eta) - \sinh 5\eta + \sinh 3\eta + \sinh \eta \,, \quad\quad
b =  \sinh (u - 3\eta) + \sinh 3\eta  \,,  \non \\
c &=& \sinh (u - 5\eta) + \sinh \eta  \,, \quad\quad\quad\quad\quad\quad\quad\quad
d =  \sinh (u - \eta) + \sinh \eta \,,  \non \\
e &=& -2 e^{-{u\over 2}} \sinh 2\eta \ \cosh ({u\over 2} - 3\eta)  \,, 
\quad\quad\quad\quad\quad
\bar e = -2 e^{{u\over 2}} \sinh 2\eta \ \cosh ({u\over 2} - 3\eta)  \,, \non \\
f &=& -2 e^{-u + 2\eta} \sinh \eta \ \sinh 2\eta -  e^{-\eta} \sinh 4\eta \,, 
\quad
\bar f = 2 e^{u - 2\eta} \sinh \eta \ \sinh 2\eta -  e^{\eta} \sinh 4\eta \,,
\non \\
g &=& 2 e^{-{u\over 2} + 2\eta} \sinh {u\over 2} \ \sinh 2\eta \,, 
\quad\quad\quad\quad\quad\quad\quad
\bar g = - 2 e^{{u\over 2} - 2\eta} \sinh {u\over 2} \ \sinh 2\eta \non \,,
\ee 
and $\eta$ is again the anisotropy parameter.

For this $R$ matrix, we find two sets of boundary quantum group 
generators, to which we refer as `type I' and `type II', following the 
classification scheme introduced by Lima-Santos \cite{LS} for the 
corresponding $K$ matrices.

\subsection{Type I}

Consider the set of generators
\be
Q_{0}(u) = \left(
\begin{array}{ccc}
    -i e^{-2 \eta}  & 0 & e^{u + \sigma}  \\
     0              & 0 & 0  \\
    e^{-u - \sigma} & 0 & i e^{2 \eta}
\end{array} \right) \,, \qquad
Q_{1} = \left(
\begin{array}{ccc}
    e^{\eta + \epsilon} & e^{\eta} & 0  \\
    e^{\eta - \sigma} & 0 & 1  \\
    0 & e^{-\sigma} & -e^{-\eta + \epsilon}
\end{array} \right) \,,
\label{chargesA22I}
\ee
where $\epsilon$ and $\sigma$ are arbitrary boundary parameters. The 
intertwining relations (\ref{intertwining}) determine a
matrix $K(u)$ with the following matrix elements:
\be
K_{11} &=& 2i e^{2\epsilon+\sigma+\eta}(e^{u}-ie^{3\eta})\cosh \eta+
2 e^{3\eta}(e^{u}+ie^{\eta})(e^{u}\cosh 2\eta - i\sinh \eta) \,, \non \\
K_{12} &=& -4 e^{u+\epsilon+\sigma+4\eta}\cosh \eta \sinh u \,, \qquad 
K_{13}  = 2i e^{u+\sigma+3\eta}(e^{u}+ie^{\eta})\sinh u \,, \non \\
K_{21} &=& -4 e^{u+\epsilon+4\eta}\cosh \eta \sinh u \,, \qquad 
K_{23}  = -4i e^{2u+\epsilon+\sigma+2\eta}\cosh \eta \sinh u \,, \non \\
K_{22} &=& 2 e^{u+2\epsilon+\sigma+4\eta}(e^{u}+ie^{-3\eta})\cosh \eta+
i e^{2\eta}(e^{u}+ie^{\eta})(e^{u} - i e^{3\eta})(e^{u} - i e^{-\eta}) \,, \non \\
K_{31} &=& 2i e^{u-\sigma+3\eta}(e^{u}+ie^{\eta})\sinh u \,, \qquad 
K_{32}  = -4i e^{2u+\epsilon+2\eta}\cosh \eta \sinh u \,, \non \\
K_{33} &=& 2i e^{2u+2\epsilon+\sigma+\eta}(e^{u}-ie^{3\eta})\cosh \eta+
2 e^{u+3\eta}(e^{u}+ie^{\eta})(\cosh 2\eta - i e^{u}\sinh \eta) \,.
\label{KA22I}
\ee 
Although this solution of the boundary Yang-Baxter equation may appear
complicated, it is considerably simpler than the one given in
\cite{LS}, to which it can be shown to be equivalent. A shift of 
$K(u)$ by $u \mapsto u + i \pi$ is also a solution, by virtue
of the periodicity $R(u+ 2i\pi)=R(u)$.

\subsection{Type II}

Consider now the set of generators
\be
Q_{0}(u) = \left(
\begin{array}{ccc}
     e^{\epsilon}  & 0 & e^{u + \sigma}  \\
     0              & 0 & 0  \\
    e^{-u - \sigma} & 0 & 0
\end{array} \right) \,, \qquad
Q_{1} = \left(
\begin{array}{ccc}
    0 & -e^{\sigma} & 0  \\
    e^{\eta} & 0 & -e^{\sigma}  \\
    0 & e^{-\eta} & 0
\end{array} \right) \,,
\label{chargesA22II}
\ee
where again $\epsilon$ and $\sigma$ are arbitrary boundary parameters. 
The intertwining relations (\ref{intertwining}) determine the
following solution of the boundary Yang-Baxter equation:
\be
K(u)=\id + 2 e^{-\epsilon} \left(
\begin{array}{ccc}
    e^{-u}\sinh \eta & 0 & e^{\sigma} \sinh u  \\
    0 & -\sinh(u-\eta) & 0 \\
    e^{-\sigma} \sinh u  & 0 & e^{u}\sinh \eta 
\end{array} \right) \,.
\label{KA22II}
\ee 
This solution is equivalent to the one found by Kim \cite{Ki}, which
is classified as type II in \cite{LS}.

\section{Discussion}\label{sec:disc}

The main results of this Letter are the expressions
(\ref{charge0A1n}), (\ref{chargesA1n}) and (\ref{chargesA22I}),
(\ref{chargesA22II}) for the boundary quantum group generators for the
cases $A^{(1)}_{n-1}$ and $A^{(2)}_{2}$, respectively; and also the
simplified expressions (\ref{KA1n}), (\ref{KA22I}), (\ref{KA22II}) for
the corresponding $K$ matrices.

It remains an open question whether, for the $A^{(1)}_{n-1}$ case, the
solution discussed here is the most general.  Indeed, for the case of
the critical $Z_{n}$-symmetric $R$ matrix \cite{KS0, Ch2} with $n=3$,
which is very similar to the $A^{(1)}_{n-1}$ $R$ matrix (\ref{RA1n})
with $n=3$, Yamada has recently found \cite{Ya} a solution of the
boundary Yang-Baxter equation with one more independent boundary
parameter.

Although a principal motivation for studying boundary quantum groups
is to find solutions of the boundary Yang-Baxter equation,
the work so far (with the exception of \cite{DN}) has not yielded new 
solutions. The main difficulty is that an independent systematic 
method of constructing the boundary quantum group generators is not yet
available.  In contrast to the bulk case \cite{Ji}, one cannot exploit
(boundary) affine Toda field theory, since appropriate classical
integrable boundary conditions are not yet known \cite{BCDR}. We hope that 
by studying the known examples of boundary quantum group generators, 
it may become possible to uncover their basic algebraic structure, and to 
find generalizations to all (non-exceptional) affine Lie algebras.

\section*{Acknowledgments}

I am grateful to Y. Yamada  for sending me a copy of his preprint.
This work was supported in part by the National Science Foundation
under Grants PHY-9870101 and PHY-0098088.

\end{document}